\documentclass[conference]{IEEEtran}
\IEEEoverridecommandlockouts 

\usepackage{amsmath, amssymb, amsfonts}

\usepackage{algorithm}
\usepackage{algpseudocode}

\usepackage{graphicx}
\usepackage{subfig}
\usepackage{booktabs}
\usepackage{stfloats}
\usepackage{array}
\usepackage{textcomp}
\usepackage{xcolor}

\usepackage{cite}
\usepackage{url}

\usepackage{verbatim}
\def\BibTeX{{\rm B\kern-.05em{\sc i\kern-.025em b}\kern-.08em
    T\kern-.1667em\lower.7ex\hbox{E}\kern-.125emX}}

\makeatletter
\newcommand{\linebreakand}{%
  \end{@IEEEauthorhalign}
  \hfill\mbox{}\par
  \mbox{}\hfill\begin{@IEEEauthorhalign}
}
\makeatother

\begin{document}

\title{DiT-SGCR: Directed Temporal Structural Representation with Global-Cluster Awareness for Ethereum Malicious Account Detection
}

\author{\IEEEauthorblockN{1\textsuperscript{st} Ye Tian}
\IEEEauthorblockA{\textit{Xidian University} \\
\textit{Hangzhou Insitute of Technology}\\
Hangzhou, China \\
tianye@xidian.edu.cn}
\and
\IEEEauthorblockN{2\textsuperscript{nd} Liangliang Song}
\IEEEauthorblockA{\textit{Xidian University} \\
\textit{Hangzhou Insitute of Technology}\\
Hangzhou, China \\
songliangl@stu.xidian.edu.cn}
\and
\IEEEauthorblockN{3\textsuperscript{rd} Peng Qian}
\IEEEauthorblockA{\textit{Zhejiang University} \\
\textit{College of Computer Science and Technology}\\
Hangzhou, China \\
messi.qp711@gmail.com}
\and
\IEEEauthorblockN{4\textsuperscript{th} Yanbin Wang$^{\ast}$ \thanks{*Yanbin Wang and Jianguo Sun are co-corresponding authors.}}
\IEEEauthorblockA{\textit{Xidian University} \\
\textit{Hangzhou Insitute of Technology}\\
Hangzhou, China \\
wangyanbin15@mails.ucas.ac.cn}
\and
\IEEEauthorblockN{5\textsuperscript{th} Jianguo Sun$^{\ast}$ }
\IEEEauthorblockA{\textit{Xidian University} \\
\textit{Hangzhou Insitute of Technology}\\
Hangzhou, China \\
jgsun@xidian.edu.cn}
\and
\IEEEauthorblockN{6\textsuperscript{th} Yifan Jia}
\IEEEauthorblockA{\textit{Harbin Engineering University} \\
\textit{Yantai Research Institute}\\
Yantai, China \\
jiayf@hrbeu.edu.cn}

}

\maketitle

\begin{abstract}
The detection of malicious accounts on Ethereum - the preeminent DeFi platform - is critical for protecting digital assets and maintaining trust in decentralized finance. Recent advances highlight that temporal transaction evolution reveals more attack signatures than static graphs. However, current methods either fail to model continuous transaction dynamics or incur high computational costs that limit scalability to large-scale transaction networks. Furthermore, current methods fail to consider two higher-order behavioral fingerprints: (1) direction in temporal transaction flows, which encodes money movement trajectories, and (2) account clustering, which reveals coordinated behavior of organized malicious collectives. To address these challenges, we propose DiT-SGCR, an unsupervised graph encoder for malicious account detection. Specifically, DiT-SGCR employs directional temporal aggregation to capture dynamic account interactions, then coupled with differentiable clustering and graph Laplacian regularization to generate high-quality, low-dimensional embeddings. Our approach simultaneously encodes directional temporal dynamics, global topology, and cluster-specific behavioral patterns, thereby enhancing the discriminability and robustness of account representations. Furthermore, DiT-SGCR  bypasses conventional graph propagation mechanisms, yielding significant scalability advantages. Extensive experiments on three datasets demonstrate that DiT-SGCR consistently outperforms state-of-the-art methods across all benchmarks, achieving F1-score improvements ranging from 3.62\% to 10.83\%. Our code is available at the following link: https://github.com/dec7l/DiT-SGCR.
\end{abstract}


\section{Introduction}
In recent years, blockchain technology has gained significant traction due to its decentralized architecture and immutability~\cite{zheng2018blockchain}. Ethereum~\cite{wood2014ethereum} has been the primary infrastructure for decentralized finance (DeFi) applications, supporting widespread adoption of smart contract-based DApps and digital assets like tokens~\cite{wust2018you,deepa2022survey}, and has spurred substantial growth in blockchain-based financial services~\cite{pal2021blockchain,iyengar2023economics,chen2024economic}. 

\begin{figure}[h!]
    \centering
    \includegraphics[width=0.85\columnwidth]{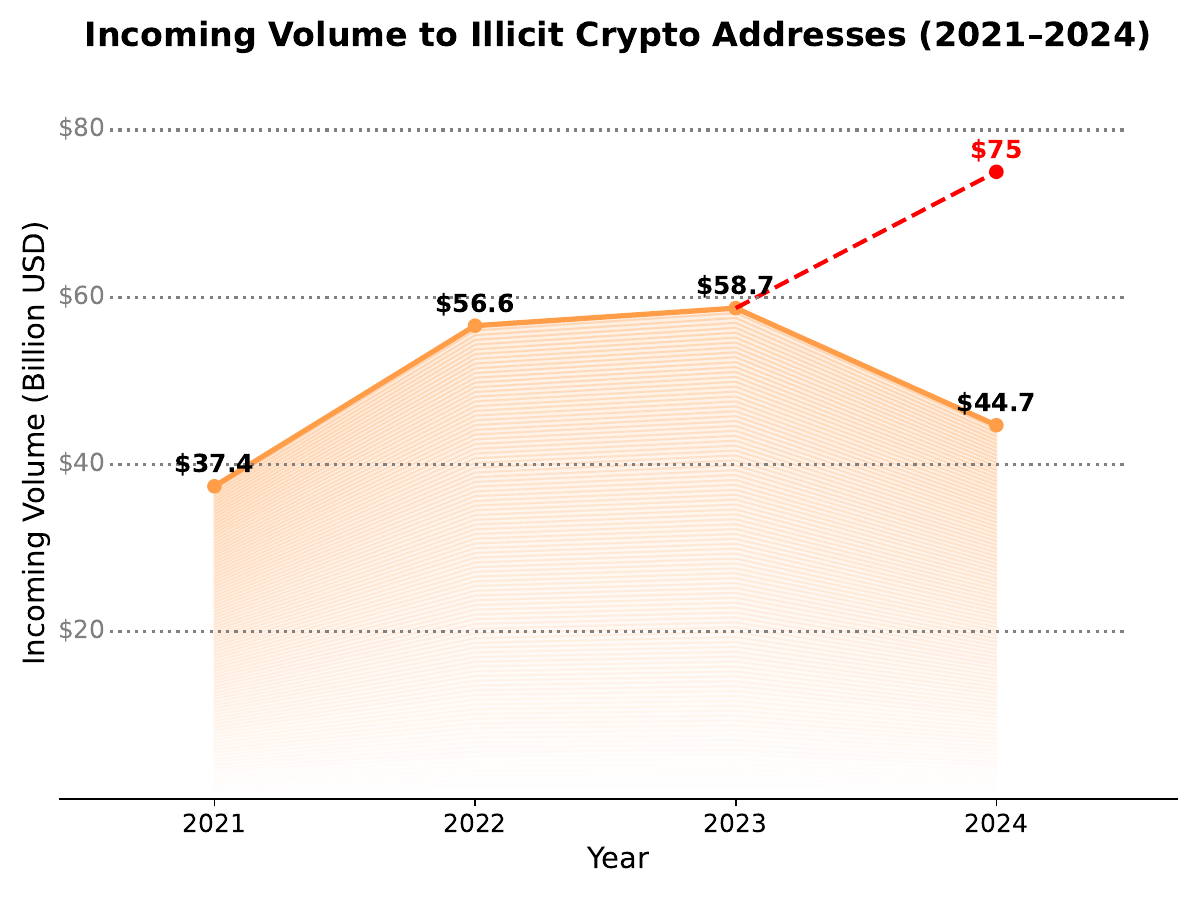}
    \caption{Incoming volume to illicit crypto address (2021-2024)}
    \label{incoming volume}
\end{figure}

However, Ethereum’s rapid adoption has been accompanied by escalating security and socioeconomic challenges \cite{chen2020survey,wang2021ethereum,chen2024dissecting}. Malicious actors increasingly exploit the pseudonymous nature of blockchain transactions to conduct fraudulent activities, undermining the platform’s integrity and stability. As shown in Figure~\ref{incoming volume}, data from TRM Labs’ 2025 Crypto Crime Report \cite{TRMLabs2025} reveals a concerning trend: illicit transaction volumes have grown from \$37.4 billion in 2021 to \$44.7 billion in 2024. Notably, due to inherent delays in blockchain forensic analysis, TRM projects the final 2024 figure may reach \$75 billion upon complete attribution - representing a potential 100\% increase from 2021 levels. This dramatic growth underscores the urgent need for effective detection mechanisms to combat evolving threats in the Ethereum ecosystem.

Historical transaction records serve as a primary data source for malicious account detection in Ethereum. These records inherently form graph-structured data, making transaction graph learning a fundamental approach for building detection models. While most prior work has focused on analyzing the structural properties of transaction graphs, recent studies highlight the critical role of temporal information in detecting malicious accounts. This is because fraudulent activities often exhibit distinctive temporal patterns—such as bursty transactions, synchronized coordination, or time-dependent money flows—that cannot be captured by structural features alone. 

Existing graph-based methods and their utilization of temporal information can be categorized as follows: a) Random Walk-Based Methods \cite{xia2019random}: These generate node embeddings to capture topological features, enabling anomaly detection via proximity analysis. While they can indirectly incorporate temporal information (e.g., through edge timestamps), they fail to explicitly model continuous transaction dynamics. 2) Static Graph Neural Networks (GNNs) \cite{zhou2020graph}: These methods aggregate topological information to learn complex transaction relationships and enable detection through neighborhood feature propagation. While they can process temporal features via snapshot-based approaches, such methods inherently lose fine-grained time resolution due to their static graph formulation. 3) Temporal Graph Neural Networks (TGNNs) \cite{longa2023graph}: These\cite{pareja2020evolvegcn,rossi2020temporal,xu2020inductive} address the above limitation by integrating time-aware mechanisms (e.g., recurrent units or attention) to track dynamic behavior. While TGNNs improve temporal modeling, they often incur high computational overhead.

In summary, existing solutions either lack explicit mechanisms to model dynamic transaction networks or require substantial resources to learn dynamic temporal features, limiting their scalability for large-scale Ethereum transaction graphs. Additionally, prior work overlooks two critical features in dynamic transaction networks:
\begin{enumerate}
\item Transaction Direction: The direction of transactions is associated with multifaceted information, such as (i) explicit tracking of fund flow trajectories, (ii) differentiation of account roles (e.g., high-outdegree sources vs. high-indegree sinks), and (iii) implicit inference of malicious intent through periodic or abrupt directional patterns.
\item Transaction Account Clustering: Clustering accounts with similar transaction behavior patterns can uncover critical insights, including anomalies, or coordinated malicious communities\cite{liu2024fishing} (e.g., phishing rings).
\end{enumerate}

To address existing challenges, we propose DiT-SGCR, an unsupervised framework that advances Ethereum fraud detection by simultaneously: (i) capturing directed temporal transaction dynamics, (ii) maintaining low computational overhead, and (iii) explicitly modeling directional flows and account clusters. First, directed temporal aggregation captures transaction directionality and temporal dynamics, generating embeddings that reflect the evolving structure of the transaction network. Next, differentiable K-means clustering derives low-dimensional structural embeddings by computing normalized distances to cluster centroids, effectively encoding coarse-grain behavioral patterns. Finally, graph Laplacian optimization refines these embeddings, balancing global topology and cluster coherence to enhance structural discriminability. DiT-SGCR provides a robust solution for detecting malicious accounts in dynamic, directed transaction networks.

Our primary contributions are summarized as follows:
\begin{itemize}
    \item Our proposed DiT-SGCR advances Ethereum security through its novel unsupervised modeling of directional transaction flows, delivering 3.62\%-10.83\% detection improvements over SOTA baselines.
    \item Our method introduce a directed temporal aggregation mechanism to capture direction-sensitive, time-evolving transaction patterns.
    \item Our method integrates a clustering mechanism to encode transactional community features, and generate robust embeddings by combining centroid distance normalization with Laplacian optimization.
    \item Our method bypasses computationally intensive graph propagation learning, making large-scale Ethereum transaction analysis computationally feasible.
    \item We release a new Ethereum phishing dataset.
\end{itemize}

\section{Related Work}
\label{sec:related_work}

Due to the dynamic, directed, and complex nature of Ethereum transaction networks, researchers have proposed various methods, ranging from traditional machine learning to temporal graph neural networks, progressively enhancing the modeling of transaction topology and temporal dynamics. This section reviews related work, describes the characteristics and representative studies of each method category, and introduces the advantages of our approach.

Traditional machine learning methods have been widely applied to detect anomalous accounts in Ethereum transaction networks by leveraging statistical features. For instance, Agarwal et al. \cite{agarwal2021detecting} proposed a framework that extracts features such as transaction frequency, balance changes, and burstiness (e.g., degree and balance bursts) from directed graphs, employing ExtraTreesClassifier for supervised classification and K-Means for unsupervised clustering to identify suspicious accounts. These approaches are computationally efficient and effective for small-scale datasets, as they rely on straightforward feature engineering and established algorithms, making them suitable for initial explorations in transaction network analysis.

Random walk-based network embedding methods focus on capturing the topological structure of transaction networks through low-dimensional node representations. DeepWalk \cite{perozzi2014deepwalk} generates node sequences via unbiased random walks and trains a Skip-Gram model to produce embeddings. Node2vec \cite{grover2016node2vec} extends this by introducing biased walks to balance depth-first and breadth-first exploration, enhancing embedding flexibility. In the Ethereum context, Wu et al. \cite{wu2020phishers} developed trans2vec, which incorporates transaction amounts and timestamps into biased random walks to generate embeddings tailored for anomaly detection. Chen et al. \cite{liu2023graph} further refined random walks by integrating gas price features for wash trading detection. These methods offer computational efficiency and scalability, effectively capturing local and global network structures in large-scale graphs.

Static graph neural networks (GNNs) have advanced anomaly detection by modeling complex topological relationships in transaction networks. Alarab et al. \cite{alarab2020competence} employed Graph Convolutional Networks (GCNs) \cite{kipf2016semi} to identify suspicious transaction patterns for anti-money laundering in Bitcoin. Graph Attention Networks (GATs) \cite{velivckovic2017graph} introduced by Veličković et al. enhance this by assigning attention weights to neighbors, prioritizing significant transaction relationships. GraphSAGE \cite{hamilton2017inductive}, proposed by Hamilton et al., uses node sampling and inductive learning to scale to large graphs, suitable for Ethereum networks. Li et al.\cite{li2023siege} proposed SIEGE, a self-supervised incremental deep graph learning model for phishing scam detection on Ethereum. Jia et al. \cite{sun2025ethereum} combined GCNs with transaction semantic features to improve fraud detection. Static GNNs excel in capturing topological patterns and support end-to-end training, making them robust for structured data analysis.

Temporal graph neural networks explicitly model the temporal dynamics of transaction networks, addressing the evolving nature of Ethereum data. Wang et al. \cite{wang2023phishing} proposed PDTGA, based on Temporal Graph Attention Network (TGAT), which uses time-encoding functions to model interactions between timestamps, node features, and edge attributes. Li et al.\cite{li2022ttagn} proposed TTAGN, a graph network model that incorporates temporal transaction information for Ethereum phishing scams detection. Grabphisher\cite{zhang2024grabphisher} models temporal and structural dynamics of accounts, representing their evolution within a continuous-time diffusion network. Wu et al. \cite{wu2024tokenscout} developed TokenScout, utilizing temporal GNNs and contrastive learning for scam token detection. These methods significantly improve the modeling of dynamic transaction patterns through temporal mechanisms.

Despite their contributions, existing methods exhibit several limitations. Traditional machine learning approaches rely on manual feature engineering, failing to capture complex topological and continuous temporal dynamics in transaction networks. Random walk-based methods model time indirectly through feature encoding, neglecting explicit temporal evolution, and overlook the directed nature of transactions, missing critical fund flow information. Static GNNs, while effective for topology, handle time via snapshots or feature encoding, losing continuous temporal information. Temporal GNNs, such as TGAT and TGN, explicitly model time, but incorporating temporal features increases computational costs and often emphasizes global network patterns over local transaction details.

To address these shortcomings, we propose DiT-SGCR, a novel directed temporal graph method that explicitly models transaction directionality and temporal dynamics through directed temporal aggregation, generates low-dimensional structural embeddings via differentiable clustering, and refines these embeddings using graph Laplacian optimization. By leveraging clustering-driven embeddings and structural optimization, DiT-SGCR effectively captures both local transaction details and global topological patterns in Ethereum transaction networks, offering superior performance in malicious account detection.

\begin{figure*}[!ht]
    \centering
    \includegraphics[width=\linewidth]{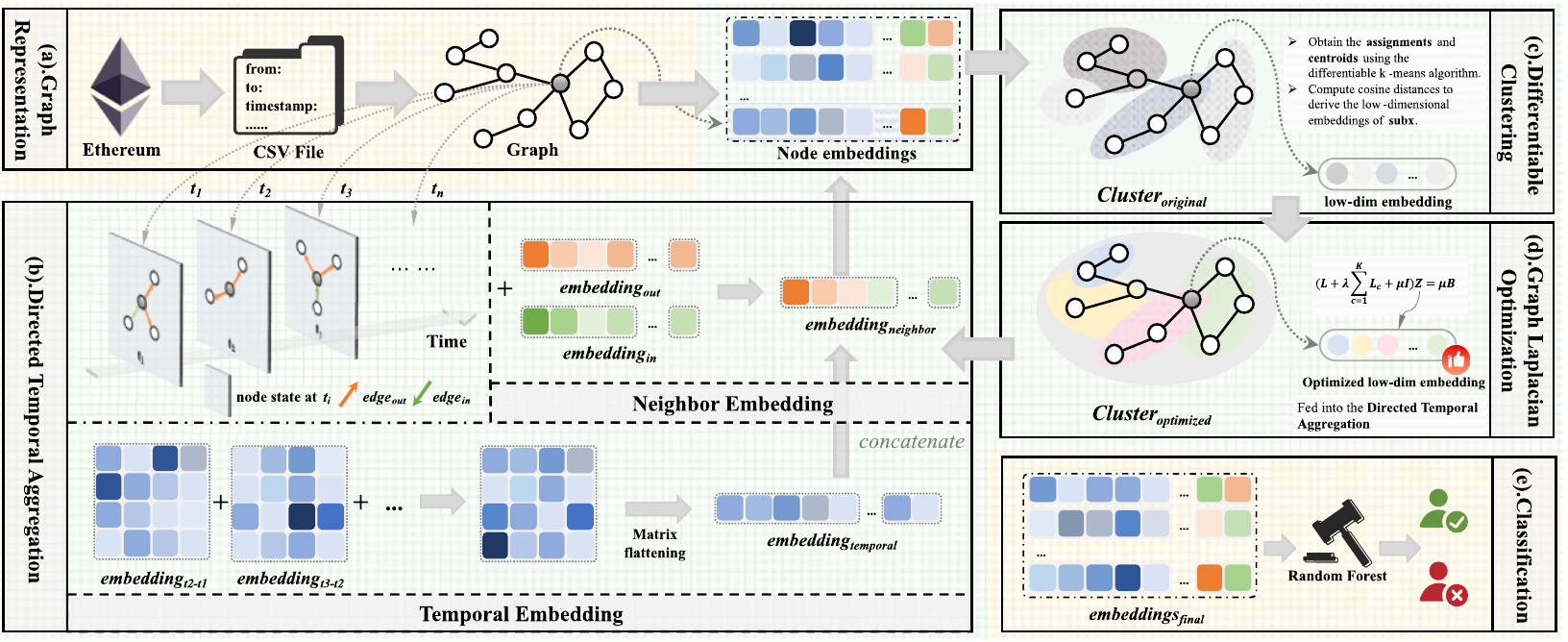}
    \caption{Architecture of the DiT-SGCR. (a) Graph Representation: Converts Ethereum transaction data from a CSV file into a directed temporal graph with nodes as accounts and edges as timestamped transactions. (b) Directed Temporal Aggregation: Aggregates embeddings from incoming and outgoing neighbors across time steps, incorporating temporal dynamics. (c) Differentiable Clustering: Applies differentiable K-means to generate low-dimensional structural embeddings. (d) Graph Laplacian Optimization: Refines embeddings by balancing global topology, cluster coherence, and fidelity to initial embeddings. (e) Classification: Uses a Random Forest classifier to detect malicious accounts based on the final embeddings. The module with the green background in the figure undergoes multiple iterations until the final embedding representation is generated, which is then passed to the yellow-highlighted process for the final classification task.}
    \label{1}
\end{figure*}

\section{Methodology}
This section elucidates the methodology of our proposed unsupervised representation learning framework for directed temporal graphs, termed \textbf{Directed Temporal Structural Global-Cluster Representation (DiT-SGCR)}, whose overall architecture is illustrated in Fig.~\ref{1}. The approach is designed to generate node embeddings that capture the evolving structural roles and temporal dynamics of nodes in directed temporal graphs. It integrates differentiable clustering to produce low-dimensional structural embeddings and employs graph Laplacian optimization to refine these embeddings, encoding both local clustering patterns and global topological structures. The overall procedure of the approach is detailed in Algorithm~\ref{alg:dit-sgcr}. The methodology is organized into the following subsections: Graph Representation, Directed Temporal Aggregation, Differentiable Clustering, Graph Laplacian Optimization, and classification.

\subsection{Graph Representation}
\label{subsec:graph-representation}

Consider a directed temporal graph \( \mathcal{G} = (V, \mathcal{E}, \mathcal{T}) \), where \( V \) denotes the set of nodes, \( \mathcal{E} \subseteq V \times V \times \mathbb{R} \) represents directed temporal edges with associated timestamps, and \( \mathcal{T} = \{t_1, t_2, \dots, t_T\} \) is the set of timestamps. Each node \( v \in V \) is associated with a sequence of interactions at timestamps \( t_i \), characterized by incoming neighbors \( \mathcal{N}_{\text{in}}(v, t_i) \) and outgoing neighbors \( \mathcal{N}_{\text{out}}(v, t_i) \). The objective is to derive a node embedding matrix \( \mathbf{H} \in \mathbb{R}^{|V| \times d} \), where \( d \) is the embedding dimension, capturing both structural and temporal characteristics of nodes.

The input data is derived from Ethereum blockchain transactions, where each node denotes an Ethereum account, including both externally owned accounts and smart contracts. Directed edges represent transactions initiated from a source account to a target account, each associated with a specific timestamp. Every transaction is defined by three key attributes: the direction of transfer, the time of occurrence, and the transferred value, typically measured in Ether. This structure inherently captures the temporal and directional characteristics of economic interactions on the Ethereum network.

In our approach, we exclusively utilize the timestamps of transactions, and the algorithm is designed to maximally exploit the temporal information encoded in these timestamps. This data is processed from a CSV file into a dictionary-based representation, where each node \( v \) stores a list of tuples \( (t_i, \mathcal{N}_{\text{in}}(v, t_i), \mathcal{N}_{\text{out}}(v, t_i)) \), ordered by descending timestamps, to facilitate efficient temporal processing and analysis of transaction patterns.

\subsection{Directed Temporal Aggregation}
\label{subsec:directed-temporal-aggregation}

The directed temporal aggregation module is designed to capture the structural roles of nodes by aggregating embeddings from both their incoming and outgoing neighbors across multiple time steps. This approach builds upon established concepts in temporal graph representation learning, such as those introduced in Temporal SIR-GN~\cite{layne2023temporal}. Specifically, for each node \( v \), the aggregation process is conducted as follows:

\begin{enumerate}
    \item \textbf{Neighbor Embedding Aggregation}: 
    For each node \( v \in V \) at timestamp \( t_i \), the embeddings of incoming and outgoing neighbors are aggregated separately:
    \begin{equation}
        \mathbf{w}_{\text{in}}(v, t_i) = \sum_{u \in \mathcal{N}_{\text{in}}(v, t_i)} \mathbf{h}_u, \quad \mathbf{w}_{\text{out}}(v, t_i) = \sum_{u \in \mathcal{N}_{\text{out}}(v, t_i)} \mathbf{h}_u,
    \end{equation}
    where \( \mathbf{h}_u \in \mathbb{R}^K \) is the initial embedding of node \( u \), and \( K \) is the number of clusters. These vectors are concatenated and normalized to unit length to form the timestamp-specific neighbor embedding:
    \begin{equation}
        \mathbf{w}(v, t_i) = \frac{[\mathbf{w}_{\text{in}}(v, t_i), \mathbf{w}_{\text{out}}(v, t_i)]}{\| [\mathbf{w}_{\text{in}}(v, t_i), \mathbf{w}_{\text{out}}(v, t_i)] \|_2 + 10^{-10}},
    \end{equation}
    where \( \mathbf{w}(v, t_i) \in \mathbb{R}^{2K} \), and the constant \( 10^{-10} \) ensures numerical stability. Additionally, a summed neighbor embedding is computed by aggregating the timestamp-specific embeddings across all timestamps:
    \begin{equation}
        \mathbf{s}_v = \sum_{t_i} \mathbf{w}(v, t_i),
    \end{equation}
    where \( \mathbf{s}_v \in \mathbb{R}^{2K} \) captures the aggregated neighbor information without temporal weighting.

    \item \textbf{Temporal Embedding Aggregation}: 
    To incorporate temporal dynamics, a structural embedding matrix \( \mathbf{Z}_v \in \mathbb{R}^{2K \times 2K} \) is computed using an exponential decay factor \( \alpha \), which modulates the influence of earlier timestamps. Initialize \( \mathbf{z}_0 = \mathbf{0} \in \mathbb{R}^{1 \times 2K} \). For each timestamp \( t_i \) (\( i \geq 2 \)), with \( t_i > t_{i-1} \):
    \begin{equation}
        \mathbf{z}_i = \exp\left( \frac{t_i - t_{i-1}}{\alpha} \right) \cdot (\mathbf{w}(v, t_{i-1})^{\top} + \mathbf{z}_{i-1}),
    \end{equation}
    followed by normalization:
    \begin{equation}
        \mathbf{z}_i = \frac{\mathbf{z}_i}{\|\mathbf{z}_i\|_2 + 10^{-10}}.
    \end{equation}
    The interaction between the current neighbor embedding \( \mathbf{w}(v, t_i) \in \mathbb{R}^{2K \times 1} \) and the temporal embedding \( \mathbf{z}_i \in \mathbb{R}^{1 \times 2K} \) is computed as an outer product:
    \begin{equation}
        \mathbf{a}_i = \mathbf{w}(v, t_i) \cdot \mathbf{z}_i,
    \end{equation}
    where \( \mathbf{a}_i \in \mathbb{R}^{2K \times 2K} \). These matrices are accumulated into:
    \begin{equation}
        \mathbf{Z}_v = \sum_{i \geq 2} \mathbf{a}_i.
    \end{equation}

    \item \textbf{Final Embedding Construction}: 
    The final embedding for node \( v \) is formed by flattening the structural embedding matrix \( \mathbf{Z}_v \) and concatenating it with the summed neighbor embedding \( \mathbf{s}_v \):
    \begin{equation}
        \mathbf{h}_v = [\text{flatten}(\mathbf{Z}_v), \mathbf{s}_v],
    \end{equation}
    where \( \text{flatten}(\mathbf{Z}_v) \in \mathbb{R}^{4K^2} \) and \( \mathbf{s}_v \in \mathbb{R}^{2K} \), resulting in a final embedding dimension of \( 4K^2 + 2K \). The embedding matrix for all nodes is \( \mathbf{H} \in \mathbb{R}^{|V| \times (4K^2 + 2K)} \), where each row corresponds to \( \mathbf{h}_v \).
\end{enumerate}

\begin{algorithm}
\caption{Process of DiT-SGCR Method}
\label{alg:dit-sgcr}
\begin{algorithmic}[1]
\Require
\State Directed temporal graph $\mathcal{G} = (V, \mathcal{E}, \mathcal{T})$
\State Number of clusters $K$
\State Temporal decay factor $\alpha$
\State Max iterations $d$
\State Inverse temperature $\beta$
\State K-means iterations $I_k$
\State Laplacian weight $\lambda$
\State Fidelity weight $\mu$
\Ensure Node embedding matrix $\mathbf{H} \in \mathbb{R}^{|V| \times (4K^2 + 2K)}$

\State Initialize $\mathbf{Z} \gets \frac{1}{K} \mathbf{1}_{|V| \times K}$
\State Load graph $\mathcal{G}$ from CSV file
\State $\mathbf{H} \gets \text{DirectedTemporalAggregation}(\mathbf{Z}, \mathcal{G}, \alpha)$
\State $\text{count} \gets \text{GetNumber}(\mathbf{H})$
\For {$i \gets 1$ to $d$}
    \State $\mathbf{H}_{\text{norm}} \gets \text{Normalize}(\mathbf{H})$
    \State $\mathbf{R}, \boldsymbol{\mu} \gets \text{DifferentiableKMeans}(\mathbf{H}_{\text{norm}}, K, \beta, I_k)$
    \State $\text{subx} \gets \text{ComputeLowDimEmbeddings}(\mathbf{H}_{\text{norm}}, \boldsymbol{\mu})$
    \State $\mathbf{Z} \gets \text{LaplacianOptimization}(\text{subx}, \mathcal{G}, \mathbf{R}, \alpha, \lambda, \mu)$
    \State $\mathbf{H}_{\text{new}} \gets \text{DirectedTemporalAggregation}(\mathbf{Z}, \mathcal{G}, \alpha)$
    \State $\text{count}_{\text{new}} \gets \text{GetNumber}(\mathbf{H}_{\text{new}})$
    \If {$\text{count} \geq \text{count}_{\text{new}}$}
        \State \textbf{break}
    \EndIf
    \State $\mathbf{H} \gets \mathbf{H}_{\text{new}}$, $\text{count} \gets \text{count}_{\text{new}}$
\EndFor
\State Store $\mathbf{H}$ to output file
\State \Return $\mathbf{H}$
\end{algorithmic}
\end{algorithm}

\subsection{Differentiable Clustering}
\label{subsec:differentiable-clustering}

To capture coarse-grained structural patterns in the Ethereum transaction network, a differentiable K-means clustering mechanism is employed, inspired by Graph InfoClust (GIC)~\cite{mavromatis2021graph}. This process generates low-dimensional structural embeddings used for subsequent optimization, taking the embedding matrix \( \mathbf{H} \in \mathbb{R}^{|V| \times (4K^2 + 2K)} \) from Section~\ref{subsec:directed-temporal-aggregation} as input. The clustering proceeds as follows:

\begin{enumerate}
    \item \textbf{Centroid Initialization}: 
    Cluster centroids \( \boldsymbol{\mu}_k \in \mathbb{R}^{4K^2 + 2K} \), for \( k = 1, \ldots, K \), are initialized using the K-means++ algorithm to ensure robust starting points.

    \item \textbf{Soft Assignment}: 
    For normalized node embeddings \( \mathbf{H} \), where each row \( \mathbf{h}_v \in \mathbb{R}^{4K^2 + 2K} \) represents the embedding of node \( v \), cosine similarities to each centroid \( \boldsymbol{\mu}_k \) are computed:
    \begin{equation}
        \text{cos\_sim}_{vk} = \frac{\mathbf{h}_v \cdot \boldsymbol{\mu}_k}{\|\mathbf{h}_v\|_2 \, \|\boldsymbol{\mu}_k\|_2}.
    \end{equation}
    Soft assignments are derived using a softmax function parameterized by an inverse temperature \( \beta \):
    \begin{equation}
        r_{vk} = \frac{\exp(\beta \cdot \text{cos\_sim}_{vk})}{\sum_{k=1}^K \exp(\beta \cdot \text{cos\_sim}_{vk})}.
    \end{equation}

    \item \textbf{Centroid Update}: 
    Centroids are recalculated as the weighted average of node embeddings:
    \begin{equation}
        \boldsymbol{\mu}_k = \frac{\sum_{v \in V} r_{vk} \mathbf{h}_v}{\sum_{v \in V} r_{vk} + 10^{-10}},
    \end{equation}
    followed by normalization to unit length.

    \item \textbf{Low-Dimensional Embedding Generation}: 
    A low-dimensional embedding matrix \( \mathbf{X} \in \mathbb{R}^{|V| \times K} \), denoted as \( \text{subx} \), is derived from the cosine distances to cluster centroids:
    \begin{equation}
        \text{val}_{vk} = 1 - \text{cos\_sim}_{vk},
    \end{equation}
    where \( \text{val}_{vk} \) is the cosine distance for node \( v \) to centroid \( k \). These distances are normalized relative to the maximum and minimum distances per node:
    \begin{equation}
        \mathbf{M}_v = \max_k (\text{val}_{vk}), \quad \mathbf{m}_v = \min_k (\text{val}_{vk}),
    \end{equation}
    \begin{equation}
        x_{vk} = \frac{\mathbf{M}_v - \text{val}_{vk}}{\mathbf{M}_v - \mathbf{m}_v + 10^{-10}}.
    \end{equation}
    The embeddings are further normalized to form a probability-like distribution:
    \begin{equation}
        \text{subx}_{vk} = \frac{x_{vk}}{\sum_{k=1}^K x_{vk} + 10^{-10}}.
    \end{equation}
    The resulting \( \text{subx} \in \mathbb{R}^{|V| \times K} \) serves as a low-dimensional structural embedding, capturing each node's proximity to cluster centroids.
\end{enumerate}

\subsection{Graph Laplacian Optimization}
\label{subsec:graph-laplacian-optimization}

To refine the low-dimensional structural embeddings for Ethereum malicious account detection, we employ a graph Laplacian optimization framework that enhances the structural coherence of the embeddings \( \text{subx} \in \mathbb{R}^{|V| \times K} \). This approach balances global topological constraints, cluster coherence, and fidelity to the initial embeddings, ensuring robust representation of node behaviors.

The optimization objective is formulated as:
\begin{equation}
    \min_{\mathbf{Z}} \text{tr}(\mathbf{Z}^\top \mathbf{L} \mathbf{Z}) + \lambda \sum_{c=1}^K \text{tr}(\mathbf{Z}^\top \mathbf{L}_c \mathbf{Z}) + \mu \|\mathbf{Z} - \mathbf{B}\|_F^2,
\end{equation}
where:
\begin{itemize}
    \item \( \mathbf{Z} \in \mathbb{R}^{|V| \times K} \) is the optimized embedding matrix.
    \item \( \mathbf{L} \in \mathbb{R}^{|V| \times |V|} \) is the graph Laplacian, defined as \( \mathbf{L} = \mathbf{D} - \mathbf{A} \), where \( \mathbf{A} \) is the adjacency matrix from the directed temporal graph, and \( \mathbf{D} \) is the degree matrix.
    \item \( \mathbf{L}_c \in \mathbb{R}^{|V| \times |V|} \) is the Laplacian for cluster \( c \), constructed from the soft assignment matrix \( \mathbf{R} \), encouraging nodes in the same cluster to have similar embeddings.
    \item \( \mathbf{B} = \mu \cdot \text{subx} \) is the initial embedding matrix, ensuring the optimized embeddings retain information from \( \text{subx} \).
    \item \( \lambda \) and \( \mu \) are hyperparameters (default 1.0) balancing cluster coherence and fidelity.
\end{itemize}

The first term, \( \text{tr}(\mathbf{Z}^\top \mathbf{L} \mathbf{Z}) \), enforces global structural consistency by minimizing the embedding differences between connected nodes. The second term, \( \sum_{c=1}^K \text{tr}(\mathbf{Z}^\top \mathbf{L}_c \mathbf{Z}) \), promotes cluster coherence by aligning embeddings within the same cluster. The third term, \( \mu \|\mathbf{Z} - \mathbf{B}\|_F^2 \), preserves the structural information encoded in \( \text{subx} \).

The optimization problem is solved using conjugate gradient descent, yielding a sparse linear system:
\begin{equation}
    (\mathbf{L} + \lambda \sum_{c=1}^K \mathbf{L}_c + \mu \mathbf{I}) \mathbf{Z} = \mu \mathbf{B},
\end{equation}
where \( \mathbf{I} \) is the identity matrix. The resulting \( \mathbf{Z} \) is a refined low-dimensional embedding matrix that effectively captures both local clustering patterns and global topological structures, enhancing the detection of malicious behaviors in Ethereum transaction networks.

The embeddings \( \mathbf{Z} \) are iteratively refined through directed temporal aggregation and clustering, ensuring that the temporal dynamics and structural roles of nodes are consistently updated throughout the learning process.

\subsection{Classification}
\label{subsec:classification}

To assess the quality of the learned node embeddings for Ethereum malicious account detection, a binary classification task is conducted using a Random Forest\cite{breiman2001random} classifier. Random Forest is an ensemble learning method that constructs multiple decision trees during training and outputs the class that is the mode of the classes predicted by individual trees. In this study, the classifier is configured with 100 trees to ensure robust performance. The dataset is split into training (80\%) and testing (20\%) sets, and the model is trained to distinguish normal accounts from malicious ones. The primary focus of our evaluation lies on the negative class's Precision, Recall, and F1-Score, as well as the overall weighted F1-Score, to assess the effectiveness of the classification results.

\section{Dataset}
\label{dataset}
In this study, we utilize three distinct datasets—MulDiGraph, B4E, and Transactions Network—to comprehensively analyze Ethereum-based transaction networks and phishing activities. These datasets, whose key attributes are summarized in Table~\ref{tab:dataset_summary}, provide critical insights into account behaviors, transaction patterns, and network structures, enabling us to investigate the dynamics of malicious activities within blockchain ecosystems.

\subsection{MulDiGraph}
The MulDiGraph dataset is publicly accessible on the XBlock platform\cite{chen2019xblock} and was made available in December 2020. It comprises a large-scale Ethereum transaction network, collected through a two-hop Breadth-First Search (BFS) from identified phishing nodes. The dataset includes 2,973,489 nodes, 13,551,303 edges, and 1,165 phishing nodes, along with detailed transaction amounts and timestamps.

\subsection{B4E}
The B4E dataset\cite{hu2023bert4eth} was gathered via Ethereum nodes over a period from January 1, 2017, to May 1, 2022. It encompasses 3,220 phishing accounts and 594,038 normal accounts, with a total of 328,261 transactions involving normal accounts. The dataset comprises four data groups, including phishing accounts, normal accounts, incoming transactions, and outgoing transactions.

\subsection{Transactions Network}
The Transactions Network dataset was constructed by our team to address gaps in the original dataset provided by Trans2Vec\cite{wu2020phishers}, which lacked comprehensive information. Starting with 1,262 confirmed malicious accounts identified in Trans2Vec, we performed a two-hop BFS via the Ethereum API to gather their transaction records. This process allowed us to build a more complete dataset. The dataset currently includes 1,262 phishing accounts, with 11,874,322 transaction edges and 2,592,316 nodes, providing a detailed view of phishing activities and network interactions within the Ethereum ecosystem.

\begin{table}[h!]
\centering
\caption{Comparative summary of dataset attributes}
\label{tab:dataset_summary}
\begin{tabular}{lccc}
\toprule
\textbf{Attribute} & \textbf{MulDiGraph} & \textbf{B4E} & \textbf{Transactions Network} \\
\midrule
\textbf{Nodes}       & 2,973,489 & 597,258  & 2,592,316 \\
\textbf{Edges}   & 13,551,303 & 1,678,901 & 11,874,322 \\
\textbf{Phisher}      & 1,165     & 3,220     & 1,262 \\
\bottomrule
\end{tabular}
\end{table}

\section{Computational Complexity Analysis}
\label{sec:Computational-complexity}

The Directed Temporal Structural Graph-Clustering Representation (DiT-SGCR) method operates on directed temporal graphs \( \mathcal{G} = (V, \mathcal{E}, \mathcal{T}) \), where \( |V| \) denotes the number of nodes, \( |\mathcal{E}| \) the number of temporal edges, and \( \mathcal{T} \) the number of timestamps. We analyze its time and space complexity assuming at most \( d \) iterations and \( k \) clusters, where \( k \) is typically small (e.g., 10). Although the algorithm includes an embedding-based convergence check that often leads to early termination, we conservatively estimate the complexity under the worst-case scenario of \( d \) full iterations.

\subsubsection{Time Complexity}

Each iteration of DiT-SGCR consists of three primary components:

\paragraph{Directed Temporal Aggregation} 
This step aggregates embeddings from temporal neighbors (via both incoming and outgoing edges), incorporating an exponential temporal decay factor \( \alpha \). It takes low-dimensional embeddings of size \( |V| \times k \) as input and produces high-dimensional embeddings of size \( |V| \times (4k^2 + 2k) \). For each edge in \( |\mathcal{E}| \), vector operations on the input embedding dimension \( k \) are applied, including additions and decay weighting \( \exp(-t/\alpha) \). The per-edge cost is \( \mathcal{O}(k) \), resulting in a total cost of \( \mathcal{O}(|\mathcal{E}| \cdot k) \).

\paragraph{Differentiable K-Means}
This component clusters nodes into \( k \) soft clusters, taking high-dimensional embeddings of size \( |V| \times (4k^2 + 2k) \) as input and producing low-dimensional embeddings of size \( |V| \times k \). In each of \( I_k \) iterations (a small constant, e.g., 10), cosine similarities between all \( |V| \) embeddings and \( k \) centroids are computed, each of dimension \( 4k^2 + 2k \approx 4k^2 \), costing \( \mathcal{O}(|V| \cdot k \cdot k^2) = \mathcal{O}(|V| \cdot k^3) \). Soft assignments and centroid updates also incur \( \mathcal{O}(|V| \cdot k^3) \), yielding a per-iteration complexity of \( \mathcal{O}(|V| \cdot k^3) \). Over \( I_k \) iterations, the total cost is \( \mathcal{O}(I_k \cdot |V| \cdot k^3) \). The low-dimensional output is derived from soft assignments, costing \( \mathcal{O}(|V| \cdot k) \).

\paragraph{Laplacian Optimization}
This stage optimizes low-dimensional embeddings of size \( |V| \times k \). It constructs a sparse adjacency matrix with temporal edge weights in \( \mathcal{O}(|\mathcal{E}|) \) time and builds \( k \) cluster-specific Laplacians, costing \( \mathcal{O}(k \cdot |\mathcal{E}|) \). Optimization is performed using a conjugate gradient solver on each of the \( k \) dimensions. Each dimension involves a sparse linear system with \( \mathcal{O}(|\mathcal{E}|) \) non-zero elements, requiring \( \mathcal{O}(\sqrt{|V|}) \) iterations, each costing \( \mathcal{O}(|\mathcal{E}|) \). Thus, per-dimension cost is \( \mathcal{O}(\sqrt{|V|} \cdot |\mathcal{E}|) \), and for \( k \) dimensions, the total is \( \mathcal{O}(k \cdot \sqrt{|V|} \cdot |\mathcal{E}|) \).

Summing the costs of the three components over \( d \) iterations, the total time complexity is:
\[
\mathcal{O}\big(d \cdot (|\mathcal{E}| \cdot k + I_k \cdot |V| \cdot k^3 + k \cdot \sqrt{|V|} \cdot |\mathcal{E}|)\big).
\]
For sparse graphs (\( |\mathcal{E}| = \mathcal{O}(|V|) \)) and small constants \( k \), \( I_k \), and \( d \), the dominant term is \( \mathcal{O}(d \cdot k \cdot \sqrt{|V|} \cdot |\mathcal{E}|) \), simplifying to:
\[
\mathcal{O}( \sqrt{|V|} \cdot |\mathcal{E}|).
\]
This demonstrates strong scalability on large temporal graphs.

\subsubsection{Space Complexity}

The memory requirements of DiT-SGCR comprise the following components:

\begin{itemize}
  \item \textbf{Embedding Storage:} Low-dimensional embeddings are of size \( |V| \times k \), and high-dimensional embeddings are of size \( |V| \times (4k^2 + 2k) \approx |V| \times 4k^2 \), giving a total of \( \mathcal{O}(|V| \cdot k^2) \).
  \item \textbf{Temporal Graph:} Storing temporal edges and timestamps (as node-wise tuples) requires \( \mathcal{O}(|\mathcal{E}| + |V| \cdot \mathcal{T}) \).
  \item \textbf{Clustering Overhead:} Soft cluster assignments require \( \mathcal{O}(|V| \cdot k) \), and centroids require \( \mathcal{O}(k \cdot k^2) = \mathcal{O}(k^3) \).
  \item \textbf{Laplacians:} Adjacency matrices and \( k \) cluster-specific Laplacians store \( \mathcal{O}(k \cdot |\mathcal{E}|) \) non-zero elements.
\end{itemize}

Thus, the total space complexity is:
\[
\mathcal{O}(|V| \cdot k^2 + k \cdot |\mathcal{E}| + k^3 + |V| \cdot \mathcal{T}).
\]
Assuming timestamp coverage is balanced, i.e., \( \mathcal{T} = \mathcal{O}(|\mathcal{E}| / |V|) \), the space requirement simplifies to:
\[
\mathcal{O}(|V| \cdot k^2 + k \cdot |\mathcal{E}| + k^3).
\]
For small \( k \), this approximates to \( \mathcal{O}(|V| + |\mathcal{E}|) \), ensuring memory efficiency.

In comparison, GNN-based methods like Temporal Graph Attention Network (TGAT) incur higher costs due to their complex architecture. TGAT uses \( L \) graph attention layers with \( H \) attention heads, processing node features of dimension \( h \). Each layer samples \( s \) temporal neighbors per node, computes attention scores costing \( \mathcal{O}(|V| \cdot s \cdot H \cdot h) \), and updates features via MLP, costing \( \mathcal{O}(|V| \cdot h^2) \), yielding a per-layer complexity of \( \mathcal{O}(|V| \cdot (s \cdot H \cdot h + h^2)) \). Over \( L \) layers and \( E \) training epochs, with edge-based loss computation (\( \mathcal{O}(|\mathcal{E}| \cdot h) \)), the time complexity is:
\[
\mathcal{O}(E \cdot L \cdot (|V| \cdot h^2 + |\mathcal{E}| \cdot h)).
\]
The space complexity includes node embeddings (\( \mathcal{O}(L \cdot |V| \cdot h) \)), graph storage (\( \mathcal{O}(|V| + |\mathcal{E}|) \)), and attention parameters (\( \mathcal{O}(L \cdot h^2) \)), resulting in:
\[
\mathcal{O}(L \cdot |V| \cdot h + |\mathcal{E}| + L \cdot h^2).
\]
Since TGAT's hidden dimension \( h \) and epoch count \( E \) are typically much larger than DiT-SGCR's cluster number \( k \) and iteration count \( d \), TGAT incurs significantly higher costs compared to DiT-SGCR's \( \mathcal{O}(d \cdot \sqrt{|V|} \cdot |\mathcal{E}|) \) time and \( \mathcal{O}(|V| + |\mathcal{E}|) \) space, making DiT-SGCR more efficient and scalable.

\section{Experiments}
In this section, we present the experimental results of the DiT-SGCR method on the datasets listed in Section ~\ref{dataset}. Our objective is to explore and address key research questions, including comparisons with common detection methods, validation of module effectiveness, and analysis of parameter configurations, to demonstrate the superior performance of our approach in detecting malicious Ethereum accounts.

\begin{itemize}
    \item \textbf{RQ1}: How does DiT-SGCR perform across different datasets, and how does it compare to other common baseline detection methods?
    \item \textbf{RQ2}: Does leveraging the directionality and temporality of transaction networks, along with adopting a Laplacian optimization strategy, enhance the performance of the detection method?
    \item \textbf{RQ3}: How do different parameter settings impact the performance of our method?
\end{itemize}

\subsection{RQ1: Comparison with Baseline Methods}

\begin{figure*}[!ht]
    \centering
    \includegraphics[width=\linewidth]{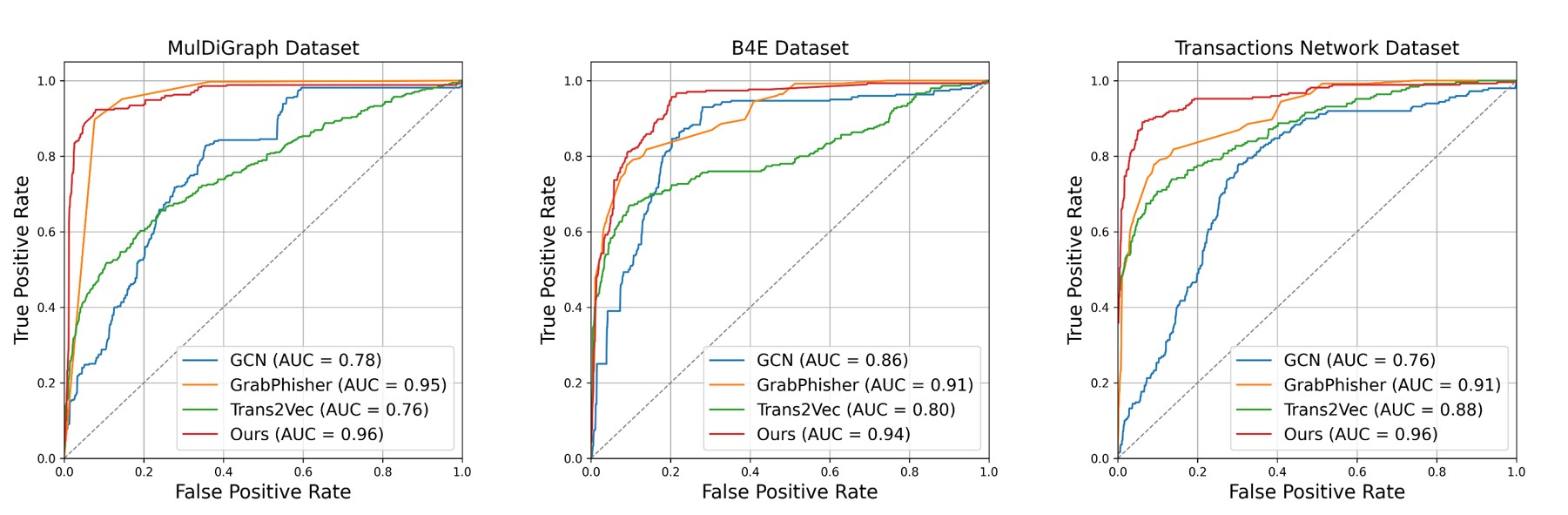}
    \caption{ROC curves of the baseline methods and DiT-SGCR on different datasets.}
    \label{roc}
\end{figure*}

In this section, we compare our proposed DiT-SGCR method with nine different baseline methods, with the experimental results presented in Table~\ref{tab:baseline_comparison} and Figure~\ref{roc}. Table~\ref{tab:baseline_comparison} provides the detailed performance comparison across all datasets, while Figure~\ref{roc} illustrates the ROC curves for representative method. To quantify the performance of our method, we use the following metrics, with their respective formulas:

\begin{table*}[h!]
\centering
\caption{Performance comparison with baseline methods on MulDiGraph, B4E, and Transactions Network datasets.}
\label{tab:baseline_comparison}
\resizebox{\textwidth}{!}{
\begin{tabular}{lcccc cccc cccc}
\toprule
\textbf{Method} & \multicolumn{4}{c}{\textbf{MulDiGraph}} & \multicolumn{4}{c}{\textbf{B4E}} & \multicolumn{4}{c}{\textbf{Transactions Network}} \\
\cmidrule(lr){2-5} \cmidrule(lr){6-9} \cmidrule(lr){10-13}
& Precision & Recall & F1-Score & W-F1 & Precision & Recall & F1-Score & W-F1 & Precision & Recall & F1-Score & W-F1 \\
\midrule
DeepWalk     & 0.1813 & 0.8293 & 0.2976 & 0.7826 & 0.4400 & 0.8919 & 0.5893 & 0.8187 & 0.5904 & 0.8963 & 0.7119 & 0.8508 \\
Role2Vec     & 0.3067 & 0.3239 & 0.3151 & 0.5583 & 0.3167 & 0.3074 & 0.3120 & 0.5328 & 0.4739 & 0.3164 & 0.3794 & 0.4685 \\
Trans2Vec    & 0.4000 & 0.8287 & 0.5396 & 0.7989 & 0.4533 & 0.9067 & 0.6044 & 0.8242 & 0.5823 & 0.8896 & 0.7039 & 0.8472 \\
\midrule
GCN          & 0.5632 & 0.6773 & 0.6150 & 0.7225 & 0.7004 & 0.6467 & 0.6724 & 0.7878 & 0.3541 & 0.9598 & 0.5173 & 0.3175 \\
GAT          & 0.6199 & 0.5307 & 0.5718 & 0.7292 & 0.6245 & 0.4767 & 0.5406 & 0.7194 & 0.5778 & 0.1044 & 0.1769 & 0.5912 \\
SAGE         & 0.6226 & 0.6027 & 0.6125 & 0.7445 & 0.8487 & 0.3367 & 0.4821 & 0.7226 & 0.8868 & 0.3775 & 0.5296 & 0.7454 \\
\midrule
TGAT         & 0.7754 & 0.8560 & 0.8137 & 0.8707 & 0.8078 & 0.7567 & 0.7814 & 0.8577 & 0.6860 & 0.9478 & 0.7960 & 0.8425 \\
GrabPhisher  & 0.8621 & 0.8974 & 0.8794 & 0.9150 & 0.8197 & 0.7549 & 0.7860 & 0.8593 & 0.8197 & 0.7549 & 0.7860 & 0.8593 \\
\midrule
BERT4ETH     & 0.5956 & 0.8901 & 0.7137 & 0.7759 & 0.5653 & 0.7441 & 0.6425 & 0.7990 & 0.6373 & 0.8112 & 0.7138 & 0.8622 \\
\midrule
\textbf{Ours} & \textbf{0.8922} & \textbf{0.9403} & \textbf{0.9156} & \textbf{0.9461} & \textbf{0.7872} & \textbf{0.8780} & \textbf{0.8301} & \textbf{0.8837} & \textbf{0.8763} & \textbf{0.9341} & \textbf{0.9043} & \textbf{0.9282} \\
\bottomrule
\end{tabular}
}
\end{table*}

\begin{itemize}
    \item \textbf{Precision}: Measures the accuracy of positive predictions, calculated as the ratio of true positives (TP) to the total number of predicted positives (TP + FP).
    \begin{equation}
        \text{Precision} = \frac{\text{TP}}{\text{TP} + \text{FP}}
    \end{equation}

    \item \textbf{Recall}: Measures the ability to identify all positive instances, calculated as the ratio of true positives (TP) to the total number of actual positives (TP + FN).
    \begin{equation}
        \text{Recall} = \frac{\text{TP}}{\text{TP} + \text{FN}}
    \end{equation}

    \item \textbf{F1-Score}: The harmonic mean of Precision and Recall, providing a single metric that balances both measures.
    \begin{equation}
        \text{F1-Score} = 2 \cdot \frac{\text{Precision} \cdot \text{Recall}}{\text{Precision} + \text{Recall}}
    \end{equation}

    \item \textbf{Weighted F1-Score (W-F1)}: Adjusts the F1-Score by weighting it according to the class distribution, addressing class imbalance. For classes \(i\) with weights \(w_i\), it is computed as:
    \begin{equation}
        \text{W-F1} = \sum_{i} w_i \cdot \text{F1-Score}_i
    \end{equation}

    \item \textbf{Area Under the Curve (AUC)\cite{fawcett2006introduction}}: Measures the overall performance of the model by calculating the area under the Receiver Operating Characteristic (ROC) curve. AUC ranges from 0 to 1, with a higher value indicating better model performance in distinguishing between positive and negative classes.
\end{itemize}

We conducted a comparative evaluation of our proposed DiT-SGCR method against multiple baseline approaches, including random walk-based methods, graph neural network based methods, temporal graph neural network-based methods, and an additional Transformer-based method. The parameters for Trans2Vec were configured based on its open-source code, while DeepWalk and Role2Vec, used as comparative baselines, employed the same parameter settings. GNN models were configured with default parameters, including 2 layers, a hidden dimension of 64, a learning rate of 0.01, and a Dropout rate of 0.5. The parameter settings for temporal graph neural networks were derived from the recommendations in the GrabPhisher paper. Similarly, the parameters for BERT4ETH were set according to the original specifications in its respective paper, ensuring fairness and consistency in the experimental setup.

To optimize the representation learning and clustering process, we set the temporal decay factor $\alpha = 1.0$ to emphasize short-term transaction patterns, and the inverse temperature $\beta = 10.0$ to balance between soft and hard cluster assignments. The number of clusters was set to $K = 10$ to effectively model underlying community structures. For Laplacian optimization, the regularization weights were set to $\lambda = 1.0$ for the cluster-specific Laplacians and $\mu = 1.0$ for the identity regularization term, jointly ensuring structural smoothness and fidelity to the initial embeddings. Early stopping was employed to terminate training once the number of unique embeddings stabilized, with a maximum of 10 iterations as a safeguard. In the final classification stage, the decision threshold was empirically set to 0.35 based on experimental performance and prior experience.

DiT-SGCR achieves superior performance, with F1-scores of 0.9156, 0.8301, and 0.9043 on the MulDiGraph, B4E, and Transactions Network datasets, respectively. In terms of AUC, DiT-SGCR attains 0.96 on both the MulDiGraph and Transactions Network datasets, and 0.94 on the B4E dataset, as illustrated in Fig \ref{roc}. Compared to the best-performing baseline, DiT-SGCR improves F1-scores by 3.62\% to 10.83\% across datasets. This substantial enhancement highlights DiT-SGCR’s capability to capture directed temporal dependencies and complex malicious behaviors, consistently outperforming state-of-the-art methods in Ethereum transaction network analysis.

In the category of random walk-based methods, we evaluate DeepWalk, Role2Vec, and Trans2Vec, with Trans2Vec being a detection model specifically proposed for Ethereum. Trans2Vec leverages timestamp and transaction amount information, achieving a notable F1-Score improvement of 0.242 on the MulDiGraph dataset compared to the other two models. Its high Recall value reflects strong detection performance; however, its Precision lags behind our method by 0.294 to 0.4922 across all datasets, resulting in a significant F1-Score gap compared to our approach. Overall, random walk-based methods struggle to effectively capture the structure of transaction networks, often exhibiting a tendency to over-identify phishing nodes.

The performance of several GNN-based methods is relatively average. Ethereum transaction networks exhibit directionality and time dependency, which are challenging for GNN methods designed for static graphs to capture effectively. During multi-layer aggregation, nodes tend to become overly similar, leading to a decline in classification capability\cite{du2021graph}; for instance, the GAT model's Recall on the Transactions Network dataset drops as low as 0.1044. In contrast, our method fully accounts for the directed and temporal features of transaction networks, fundamentally enhancing its ability to understand graph structures.

We explored temporal graph neural network methods. TGAT, by modeling temporal signals, effectively captures periodic patterns; however, its overemphasis on temporal features often leads to neglecting global graph information, resulting in reduced accuracy (e.g., a Precision of only 0.6860 on the Transactions Network dataset). GrabPhisher\cite{zhang2024grabphisher}, an improvement over TGAT, excels at capturing the dynamic evolution of graphs, demonstrating strong performance. Nevertheless, its overall approach is relatively complex and cumbersome, and its performance still falls short of our DiT-SGCR method.

We also experimented with the Transformer-based\cite{vaswani2017attention} BERT4ETH method, which demonstrates strong capabilities in sequential transaction analysis but is limited in capturing graph topological structures. Additionally, BERT4ETH requires complex preprocessing for its input, and its pretraining process is both time-consuming and heavily dependent on the practitioner's expertise\cite{bergstra2012random}, limiting its practicality in many scenarios. While BERT4ETH achieves a Recall comparable to our method, its overall performance, particularly in terms of Precision and F1-Score, remains inferior to our DiT-SGCR approach.

\subsection{RQ2: Ablation Analysis}

To evaluate the contribution of each key component in DiT-SGCR, we conducted an ablation study by systematically removing the neighbor embeddings (w/o Neighbor), temporal embeddings (w/o Temporal), and Laplacian optimization (w/o Laplacian). The experimental setup and evaluation metrics, consistent with those described in RQ1, were applied to the MultiDiGraph, B4E, and Transactions Network datasets. Performance is reported in terms of Precision, Recall, F1-Score, and W-F1 as shown in Table~\ref{tab:ablation_study}.

\begin{table*}[th!]
\centering
\caption{Ablation study: performance comparison on MulDiGraph, B4E, and Transactions Network datasets.}
\label{tab:ablation_study}
\resizebox{\textwidth}{!}{
\begin{tabular}{lcccc cccc cccc}
\toprule
\textbf{Method} & \multicolumn{4}{c}{\textbf{MulDiGraph}} & \multicolumn{4}{c}{\textbf{B4E}} & \multicolumn{4}{c}{\textbf{Transactions Network}} \\
\cmidrule(lr){2-5} \cmidrule(lr){6-9} \cmidrule(lr){10-13}
& Precision & Recall & F1-Score & W-F1 & Precision & Recall & F1-Score & W-F1 & Precision & Recall & F1-Score & W-F1 \\
\midrule
w/o Neighbor & 0.8736 & 0.8636 & 0.8686 & 0.9189 & 0.7416 & 0.8271 & 0.7821 & 0.8508 & 0.8123 & 0.8242 & 0.8182 & 0.8663 \\
w/o Temperal & 0.8720 & 0.9290 & 0.8996 & 0.9356 & 0.7588 & 0.8746 & 0.8126 & 0.8699 & 0.8556 & 0.8901 & 0.8725 & 0.9053 \\
w/o Laplacian & 0.8714 & 0.9432 & 0.9059 & 0.9392 & 0.7571 & 0.9085 & 0.8259 & 0.8769 & 0.8660 & 0.9231 & 0.8936 & 0.9202 \\
\midrule
\textbf{Ours(Full)}    & \textbf{0.8922} & \textbf{0.9403} & \textbf{0.9156} & \textbf{0.9461} & \textbf{0.7872} & \textbf{0.8780} & \textbf{0.8301} & \textbf{0.8837} & \textbf{0.8763} & \textbf{0.9341} & \textbf{0.9043} & \textbf{0.9282} \\

\bottomrule
\end{tabular}
}
\end{table*}

The complete approach achieves superior performance across all datasets, with F1-Scores of 0.9156, 0.8301, and 0.9043, and W-F1 scores of 0.9461, 0.8837, and 0.9282 on MultiDiGraph, B4E, and Transactions Network, respectively. Removing neighbor embeddings results in a significant performance decline, with F1-Scores dropping to 0.8686, 0.7821, and 0.8182, representing reductions of 4.7\%, 4.8\%, and 8.6\% compared to the full model. This substantial impact underscores the critical role of neighbor embeddings, which form the foundational information for constructing the transaction network. Their absence leads to a marked decrease in the model's ability to capture network structure, severely impairing overall performance.

Excluding temporal embeddings reduces F1-Scores to 0.8996, 0.8126, and 0.8901, corresponding to declines of 1.6\%, 1.8\%, and 1.4\%, respectively. In the context of Ethereum, temporal dynamics are highly significant, as short-term, high-frequency transactions are often indicative of malicious activities. The removal of temporal embeddings diminishes the model's ability to detect such malicious accounts, contributing to the observed performance degradation.

The absence of Laplacian optimization leads to F1-Scores of 0.9059, 0.8259, and 0.9231, with Precision dropping from 0.8922, 0.7872, and 0.8763 in the full model to 0.8714, 0.7571, and 0.8660, respectively—an average improvement of approximately 2.5\% in Precision with Laplacian optimization. This component harmonizes global and cluster-specific information, yielding embeddings that are not only robust and discriminative but also elegantly aligned with the underlying graph topology. Its inclusion significantly enhances the model's overall effectiveness, demonstrating its pivotal role in achieving state-of-the-art performance.

This ablation study validates the indispensability of neighbor embeddings, temporal embeddings, and Laplacian optimization. Their synergistic integration empowers our model to excel in directed temporal graph tasks, particularly in complex scenarios like Ethereum transaction analysis.

\subsection{RQ3: Hyper Parametric Analysis}

\begin{figure*}[!ht]
    \centering
    \includegraphics[width=\linewidth]{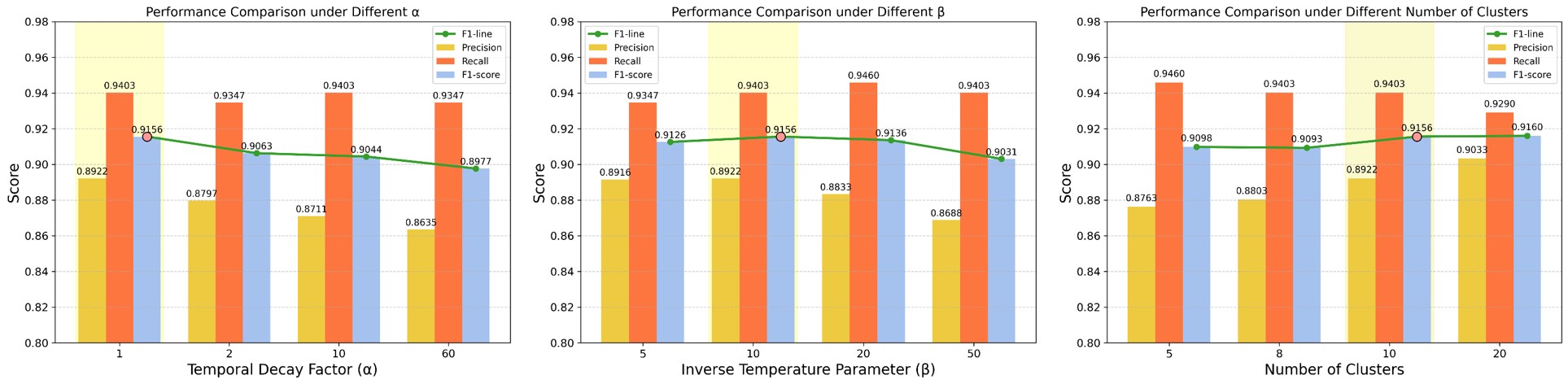}
    \caption{Bar chart showing the effect of hyperparameter variations on DiT-SGCR performance.}
    \label{barchart}
\end{figure*}

\begin{figure}[htbp]
    \centering
    \includegraphics[width=\linewidth]{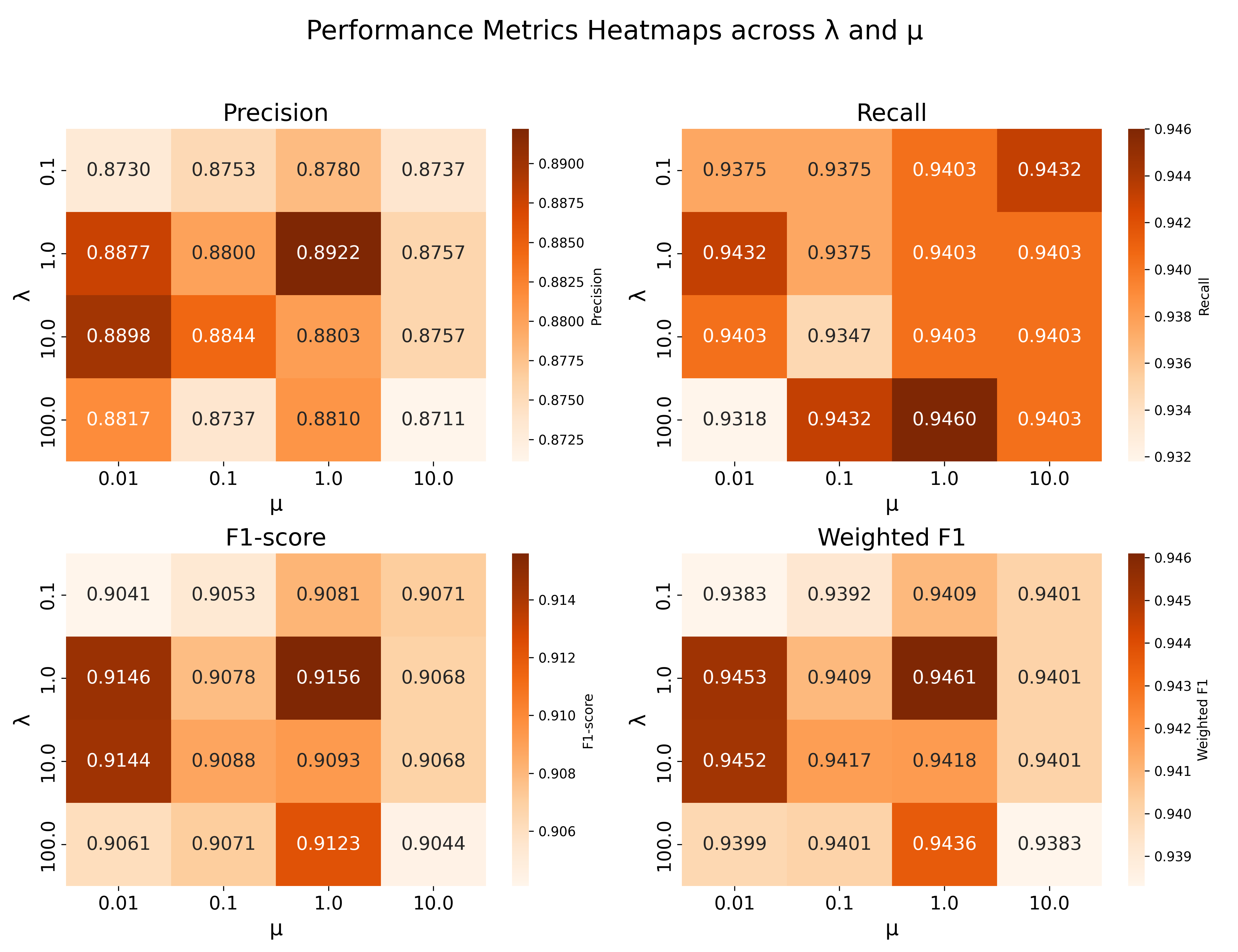}
    \caption{Heatmap showing the effect of hyperparameter variations on DiT-SGCR performance.}
    \label{heatmaps}
\end{figure}

To address Research Question RQ3, we systematically evaluated the impact of key hyperparameters in the DiT-SGCR method, specifically the temporal decay factor $\alpha$, the inverse temperature parameter $\beta$, the number of clusters $K$, and the Laplacian optimization parameters $\lambda$ and $\mu$, on embedding quality and downstream classification performance. This study utilizes the MulDiGraph dataset. The experimental results are presented in Fig.~\ref{barchart} and Fig.~\ref{heatmaps}.
 In the default configuration, we set $\alpha = 1.0$ to weigh recent and historical interactions, $\beta = 10.0$ to control soft K-means clustering assignments, $K = 10$ to capture community structures, and $\lambda = 1.0$ and $\mu = 1.0$ to balance cluster-specific Laplacian weights and initial embedding regularization. To enhance computational efficiency, we employed an early stopping mechanism based on the convergence of unique embedding counts, with a maximum of 10 iterations as a fallback for non-converging cases.

\subsubsection{Impact of Temporal Decay Factor $\alpha$}
In DiT-SGCR, the temporal decay factor $\alpha$ adjusts the influence of transaction timestamps on edge weights, enabling the construction of a dynamic adjacency matrix that captures evolving transaction behaviors. Since Ethereum logs activity in seconds, malicious accounts—especially phishing ones—often show bursty behavior in short periods. We evaluated $\alpha$ values of 1, 2, 10, and 60. Notably, $\alpha=1$ delivered the best F1 score of 0.9156, while increasing $\alpha$ to 60 reduced the F1 to 0.8977. This suggests that focusing on second-level granularity sharpens the detection of short-term anomalies, whereas a broader window introduces noise. Hence, we select \textbf{$\alpha=1$} for optimal temporal sensitivity.

\subsubsection{Effect of Inverse Temperature Parameter $\beta$}
The inverse temperature $\beta$ in differentiable K-means controls the sharpness of cluster assignments. A higher $\beta$ leads to crisper assignments, while a lower one allows softer groupings—affecting both embedding diversity and classification. We tested $\beta$ values of 5, 10, 20, and 50. The best performance emerged at \textbf{$\beta=10$} with an F1 score of 0.9156, balancing clarity and flexibility in clustering. In contrast, $\beta=50$ yielded a lower F1 of 0.9031, showing that overly hard assignments can harm discriminability. Therefore, we adopt \textbf{$\beta=10$} to ensure robust yet expressive embeddings.

\subsubsection{Influence of Cluster Count}
The \texttt{clusters} parameter determines how embeddings are partitioned in K-means, directly influencing model complexity and performance. We explored values of 5, 8, 10, and 20. Performance improved with more clusters, but at \texttt{clusters=20}, computational cost increased significantly. \texttt{clusters=10} emerged as the sweet spot—offering high detection accuracy with efficient runtime—making it our choice for the final model.

\subsubsection{Optimization Weights $\lambda$ and $\mu$}
The hyperparameters $\lambda$ and $\mu$ steer the optimization of the cluster-specific Laplacian and embedding alignment. We experimented with $\lambda \in \{0.1, 1, 10, 100\}$ and $\mu \in \{0.01, 0.1, 1, 10\}$. Peak performance was achieved at $\lambda=1.0$ and $\mu=1.0$, with an F1 score of 0.9461. Performance remained strong across moderate values but declined when these weights were too high, indicating over-regularization. As such, we choose \textbf{$\lambda=1.0$} and \textbf{$\mu=1.0$} to balance structure preservation with embedding fidelity.

\section{Limitations}
\label{sec:limitations}

While DiT-SGCR achieves significant improvements in Ethereum fraud detection, several limitations persist that should be acknowledged and addressed in future research.

\textbf{Feature Set Restriction}: DiT-SGCR relies solely on timestamps for graph construction due to current algorithmic constraints. Incorporating richer transactional features—such as transferred amounts and Gas fees—could improve embedding quality and enhance the model's ability to detect phishing activities.

\textbf{Static Graph Processing}: DiT-SGCR assumes a static transaction graph, which limits its applicability in real-world scenarios. Enabling incremental graph updates would allow the model to adapt to new transactions in real time, supporting timely fraud detection and improving scalability for large-scale dynamic environments.

\textbf{Dataset Temporal Coverage}: Although we provide a new dataset, it is still based on historical malicious accounts and thus remains somewhat outdated. Moreover, acquiring up-to-date scam addresses via the Ethereum explorer poses practical challenges. Longitudinal datasets incorporating recent fraud patterns could enhance the generalizability and robustness of evaluation.

\textbf{Blockchain-Specific Testing}: Current research, including our own, primarily focuses on the Ethereum platform. Generalizing the approach to Bitcoin or other blockchain systems could broaden its applicability across diverse financial security scenarios.

\section{Conclusion}
\label{sec:conclusion}

In this paper, we introduced DiT-SGCR, a novel clustering-driven framework for capturing structural and temporal representations of Ethereum transaction networks aimed at malicious account detection. Our method uniquely integrates directional temporal aggregation to model dynamic money flows, differentiable clustering to reveal coordinated behaviors among accounts, and graph Laplacian regularization to preserve global structural properties within low-dimensional embeddings. By bypassing conventional graph propagation mechanisms, DiT-SGCR achieves superior scalability, making it well-suited for large-scale blockchain environments. Extensive experiments across three real-world datasets demonstrate that DiT-SGCR consistently outperforms state-of-the-art baselines, achieving F1-score improvements ranging from 3.62\% to 10.83\%, thereby validating its robustness and effectiveness in detecting fraud on decentralized platforms. These findings underscore the importance of capturing directional temporal dynamics and cluster-specific behavioral patterns in advancing blockchain security analytics. As part of future work, we plan to incorporate richer transaction features (e.g., transferred amounts, Gas fees), develop incremental graph processing capabilities to support real-time updates, utilize longitudinal datasets with recent fraud patterns to improve generalizability, and extend our framework to other blockchain platforms such as Bitcoin, broadening its applicability in diverse financial security scenarios.

\bibliographystyle{elsarticle-num} 
\bibliography{my_ref}

\begin{thebibliography}{10}
\expandafter\ifx\csname url\endcsname\relax
  \def\url#1{\texttt{#1}}\fi
\expandafter\ifx\csname urlprefix\endcsname\relax\def\urlprefix{URL }\fi
\expandafter\ifx\csname href\endcsname\relax
  \def\href#1#2{#2} \def\path#1{#1}\fi

\bibitem{zheng2018blockchain}
Z.~Zheng, S.~Xie, H.-N. Dai, X.~Chen, H.~Wang, Blockchain challenges and opportunities: A survey, International journal of web and grid services 14~(4) (2018) 352--375.

\bibitem{wood2014ethereum}
G.~Wood, et~al., Ethereum: A secure decentralised generalised transaction ledger, Ethereum project yellow paper 151~(2014) (2014) 1--32.

\bibitem{wust2018you}
K.~W{\"u}st, A.~Gervais, Do you need a blockchain?, in: 2018 crypto valley conference on blockchain technology (CVCBT), IEEE, 2018, pp. 45--54.

\bibitem{deepa2022survey}
N.~Deepa, Q.-V. Pham, D.~C. Nguyen, S.~Bhattacharya, B.~Prabadevi, T.~R. Gadekallu, P.~K.~R. Maddikunta, F.~Fang, P.~N. Pathirana, A survey on blockchain for big data: Approaches, opportunities, and future directions, Future Generation Computer Systems 131 (2022) 209--226.

\bibitem{pal2021blockchain}
A.~Pal, C.~K. Tiwari, A.~Behl, Blockchain technology in financial services: a comprehensive review of the literature, Journal of Global Operations and Strategic Sourcing 14~(1) (2021) 61--80.

\bibitem{iyengar2023economics}
G.~Iyengar, F.~Saleh, J.~Sethuraman, W.~Wang, Economics of permissioned blockchain adoption, Management Science 69~(6) (2023) 3415--3436.

\bibitem{chen2024economic}
X.~Chen, Q.~Cheng, T.~Luo, The economic value of blockchain applications: Early evidence from asset-backed securities, Management Science 70~(1) (2024) 439--463.

\bibitem{chen2020survey}
H.~Chen, M.~Pendleton, L.~Njilla, S.~Xu, A survey on ethereum systems security: Vulnerabilities, attacks, and defenses, ACM Computing Surveys (CSUR) 53~(3) (2020) 1--43.

\bibitem{wang2021ethereum}
Z.~Wang, H.~Jin, W.~Dai, K.-K.~R. Choo, D.~Zou, Ethereum smart contract security research: survey and future research opportunities, Frontiers of Computer Science 15 (2021) 1--18.

\bibitem{chen2024dissecting}
Z.~Chen, Y.~Hu, B.~He, D.~Luo, L.~Wu, Y.~Zhou, Dissecting payload-based transaction phishing on ethereum, arXiv preprint arXiv:2409.02386 (2024).

\bibitem{TRMLabs2025}
{TRM Labs}, \href{https://www.trmlabs.com/resources/reports/2025-crypto-crime-report}{2025 crypto crime report: Key trends that shaped the illicit crypto market in 2024}, accessed: 27 April 2025 (2025).
\newline\urlprefix\url{https://www.trmlabs.com/resources/reports/2025-crypto-crime-report}

\bibitem{xia2019random}
F.~Xia, J.~Liu, H.~Nie, Y.~Fu, L.~Wan, X.~Kong, Random walks: A review of algorithms and applications, IEEE Transactions on Emerging Topics in Computational Intelligence 4~(2) (2019) 95--107.

\bibitem{zhou2020graph}
J.~Zhou, G.~Cui, S.~Hu, Z.~Zhang, C.~Yang, Z.~Liu, L.~Wang, C.~Li, M.~Sun, Graph neural networks: A review of methods and applications, AI open 1 (2020) 57--81.

\bibitem{longa2023graph}
A.~Longa, V.~Lachi, G.~Santin, M.~Bianchini, B.~Lepri, P.~Lio, F.~Scarselli, A.~Passerini, Graph neural networks for temporal graphs: State of the art, open challenges, and opportunities, arXiv preprint arXiv:2302.01018 (2023).

\bibitem{pareja2020evolvegcn}
A.~Pareja, G.~Domeniconi, J.~Chen, T.~Ma, T.~Suzumura, H.~Kanezashi, T.~Kaler, T.~Schardl, C.~Leiserson, Evolvegcn: Evolving graph convolutional networks for dynamic graphs, in: Proceedings of the AAAI conference on artificial intelligence, Vol.~34, 2020, pp. 5363--5370.

\bibitem{rossi2020temporal}
E.~Rossi, B.~Chamberlain, F.~Frasca, D.~Eynard, F.~Monti, M.~Bronstein, Temporal graph networks for deep learning on dynamic graphs, arXiv preprint arXiv:2006.10637 (2020).

\bibitem{xu2020inductive}
D.~Xu, C.~Ruan, E.~Korpeoglu, S.~Kumar, K.~Achan, Inductive representation learning on temporal graphs, arXiv preprint arXiv:2002.07962 (2020).

\bibitem{liu2024fishing}
J.~Liu, J.~Chen, J.~Wu, Z.~Wu, J.~Fang, Z.~Zheng, Fishing for fraudsters: Uncovering ethereum phishing gangs with blockchain data, IEEE Transactions on Information Forensics and Security 19 (2024) 3038--3050.

\bibitem{agarwal2021detecting}
R.~Agarwal, S.~Barve, S.~K. Shukla, Detecting malicious accounts in permissionless blockchains using temporal graph properties, Applied Network Science 6 (2021) 1--30.

\bibitem{perozzi2014deepwalk}
B.~Perozzi, R.~Al-Rfou, S.~Skiena, Deepwalk: Online learning of social representations, in: Proceedings of the 20th ACM SIGKDD international conference on Knowledge discovery and data mining, 2014, pp. 701--710.

\bibitem{grover2016node2vec}
A.~Grover, J.~Leskovec, node2vec: Scalable feature learning for networks, in: Proceedings of the 22nd ACM SIGKDD international conference on Knowledge discovery and data mining, 2016, pp. 855--864.

\bibitem{wu2020phishers}
J.~Wu, Q.~Yuan, D.~Lin, W.~You, W.~Chen, C.~Chen, Z.~Zheng, Who are the phishers? phishing scam detection on ethereum via network embedding, IEEE Transactions on Systems, Man, and Cybernetics: Systems 52~(2) (2020) 1156--1166.

\bibitem{liu2023graph}
J.~Liu, C.~Yin, H.~Wang, X.~Wu, D.~Lan, L.~Zhou, C.~Ge, Graph embedding-based money laundering detection for ethereum, Electronics 12~(14) (2023) 3180.

\bibitem{alarab2020competence}
I.~Alarab, S.~Prakoonwit, M.~I. Nacer, Competence of graph convolutional networks for anti-money laundering in bitcoin blockchain, in: Proceedings of the 2020 5th international conference on machine learning technologies, 2020, pp. 23--27.

\bibitem{kipf2016semi}
T.~N. Kipf, M.~Welling, Semi-supervised classification with graph convolutional networks, arXiv preprint arXiv:1609.02907 (2016).

\bibitem{velivckovic2017graph}
P.~Veli{\v{c}}kovi{\'c}, G.~Cucurull, A.~Casanova, A.~Romero, P.~Lio, Y.~Bengio, Graph attention networks, arXiv preprint arXiv:1710.10903 (2017).

\bibitem{hamilton2017inductive}
W.~Hamilton, Z.~Ying, J.~Leskovec, Inductive representation learning on large graphs, Advances in neural information processing systems 30 (2017).

\bibitem{li2023siege}
S.~Li, R.~Wang, H.~Wu, S.~Zhong, F.~Xu, Siege: Self-supervised incremental deep graph learning for ethereum phishing scam detection, in: Proceedings of the 31st ACM International Conference on Multimedia, 2023, pp. 8881--8890.

\bibitem{sun2025ethereum}
J.~Sun, Y.~Jia, Y.~Wang, Y.~Tian, S.~Zhang, Ethereum fraud detection via joint transaction language model and graph representation learning, Information Fusion 120 (2025) 103074.

\bibitem{wang2023phishing}
L.~Wang, M.~Xu, H.~Cheng, Phishing scams detection via temporal graph attention network in ethereum, Information Processing \& Management 60~(4) (2023) 103412.

\bibitem{li2022ttagn}
S.~Li, G.~Gou, C.~Liu, C.~Hou, Z.~Li, G.~Xiong, Ttagn: Temporal transaction aggregation graph network for ethereum phishing scams detection, in: Proceedings of the ACM Web Conference 2022, 2022, pp. 661--669.

\bibitem{zhang2024grabphisher}
J.~Zhang, H.~Sui, X.~Sun, C.~Ge, L.~Zhou, W.~Susilo, Grabphisher: Phishing scams detection in ethereum via temporally evolving gnns, IEEE Transactions on Services Computing (2024).

\bibitem{wu2024tokenscout}
C.~Wu, J.~Chen, Z.~Zhao, K.~He, G.~Xu, Y.~Wu, H.~Wang, H.~Li, Y.~Liu, Y.~Xiang, Tokenscout: Early detection of ethereum scam tokens via temporal graph learning, in: Proceedings of the 2024 on ACM SIGSAC Conference on Computer and Communications Security, 2024, pp. 956--970.

\bibitem{layne2023temporal}
J.~Layne, J.~Carpenter, E.~Serra, F.~Gullo, Temporal sir-gn: Efficient and effective structural representation learning for temporal graphs, Proceedings of the VLDB Endowment 16~(9) (2023) 2075--2089.

\bibitem{mavromatis2021graph}
C.~Mavromatis, G.~Karypis, Graph infoclust: Maximizing coarse-grain mutual information in graphs, in: Pacific-Asia conference on knowledge discovery and data mining, Springer, 2021, pp. 541--553.

\bibitem{breiman2001random}
L.~Breiman, Random forests, Machine learning 45 (2001) 5--32.

\bibitem{chen2019xblock}
L.~Chen, J.~Peng, Y.~Liu, J.~Li, F.~Xie, Z.~Zheng, Xblock blockchain datasets: Inpluslab ethereum phishing detection datasets (2019).

\bibitem{hu2023bert4eth}
S.~Hu, Z.~Zhang, B.~Luo, S.~Lu, B.~He, L.~Liu, Bert4eth: A pre-trained transformer for ethereum fraud detection, in: Proceedings of the ACM Web Conference 2023, 2023, pp. 2189--2197.

\bibitem{fawcett2006introduction}
T.~Fawcett, An introduction to roc analysis, Pattern recognition letters 27~(8) (2006) 861--874.

\bibitem{du2021graph}
G.~Du, J.~Zhang, M.~Jiang, J.~Long, Y.~Lin, S.~Li, K.~C. Tan, Graph-based class-imbalance learning with label enhancement, IEEE transactions on neural networks and learning systems 34~(9) (2021) 6081--6095.

\bibitem{vaswani2017attention}
A.~Vaswani, N.~Shazeer, N.~Parmar, J.~Uszkoreit, L.~Jones, A.~N. Gomez, {\L}.~Kaiser, I.~Polosukhin, Attention is all you need, Advances in neural information processing systems 30 (2017).

\bibitem{bergstra2012random}
J.~Bergstra, Y.~Bengio, Random search for hyper-parameter optimization, The journal of machine learning research 13~(1) (2012) 281--305.

\end{thebibliography}

\end{document}